\documentclass[preprint]{ptptex}
\usepackage{wrapft}
\usepackage{graphicx}

\newcommand{\bra}[1]{\langle {#1} |}     
\newcommand{\ket}[1]{| {#1} \rangle}     
\newcommand{\bbra}[1]{\langle\!\langle {#1} |}     
\newcommand{\kket}[1]{| {#1} \rangle\!\rangle}     
\newcommand{\dbra}[1]{\langle {#1} |\!|}     
\newcommand{\dket}[1]{|\!| {#1} \rangle}     
\newcommand{\maru}[1]{\stackrel{\tiny\circ} {#1}} 
\newcommand{\wtilde}[1]{\widetilde{#1}} 

\newcommand{\kbar}{k \kern -0.5em\raise 0.6ex \hbox{--}}

\def\<{\langle}
\def\>{\rangle}
\def\bsub{\begin{subequations}}
\def\esub{\end{subequations}}
\def\beqn{\begin{eqnarray}}
\def\eeqn{\end{eqnarray}}
\def\b{\begin{equation}}

\title{
A Note on the Two-Level Pairing Model Obeying the $su(2)\otimes su(2)$-Algebra
}
\subtitle{
Re-formation in Terms of the $su(1,1)\otimes su(1,1)$-Algebra
}    

\author{
Constan\c{c}a {\sc Provid\^encia},$^{1}$
Jo\~ao da {\sc Provid\^encia},$^{1}$\\
Yasuhiko {\sc Tsue}$^{2}$ 
and Masatoshi {\sc Yamamura}$^{3}$
}

\inst{
$^{1}$Departamento de Fisica, Universidade de Coimbra, 3004-516 Coimbra, 
Portugal\\
$^2$Physics Division, Faculty of Science, Kochi University, Kochi 780-8520, 
Japan\\
$^{3}$Faculty of Engineering, Kansai University, Suita 564-8680, Japan
}

\recdate{
\today
}

\abst{
The two-level pairing model obeying the $su(2)\otimes su(2)$-algebra, 
which was discussed in the previous paper, is re-formed in the framework 
of the $su(1,1)\otimes su(1,1)$-algebra in the Schwinger boson 
representation. 
With the aid of MYT mapping method, the Schwinger boson representation is 
reduced to the Holstein-Primakoff boson representation, 
which is expressed in terms of two kinds of bosons. 
Further, with the aid of the squeezed coherent state, the classical 
counterpart is obtained. 
}
\begin{document}
\maketitle


In the previous paper (referred to as (A)), the two-level pairing model in 
many-fermion system was formulated in the Schwinger boson 
representation.\cite{1}
Basic idea is the use of four kinds of boson operators. 
In (A), noting the existence of three constants of motion, 
the model is re-formed in the framework of one kind of boson. 
In the present paper, we re-form the model in terms of two kinds of bosons. 
This re-formation is closely related to the $su(1,1)\otimes su(1,1)$-algebra. 
We will not repeat the interpretation of the notations 
which already appeared in (A).

First, we introduce the $su(1,1)\otimes su(1,1)$-algebra, the generators 
of which are given as 
\bsub\label{1}
\beqn
& &{\wtilde T}_+(+)=\hbar{\hat a}_+^*{\hat b}_-^* \ , \quad
{\wtilde T}_-(+)=\hbar{\hat b}_-{\hat a}_+ \ , \quad
{\wtilde T}_0(+)=(\hbar/2)({\hat a}_+^*{\hat a}_+ + {\hat b}_-^*{\hat b}_-
+1) \ , 
\label{1a}\\
& &{\wtilde T}_+(-)=\hbar{\hat a}_-^*{\hat b}_+^* \ , \quad
{\wtilde T}_-(-)=\hbar{\hat b}_+{\hat a}_- \ , \quad
{\wtilde T}_0(-)=(\hbar/2)({\hat a}_-^*{\hat a}_- + {\hat b}_+^*{\hat b}_+
+1) \ . 
\label{1b}
\eeqn
\esub
The form (\ref{1a}) is identical to that given in the relation 
(A$\cdot$3$\cdot$5a). 
The set\break
$({\wtilde T}_{\pm,0}(+), {\wtilde T}_{\pm,0}(-))$ obeys the 
$su(1,1)\otimes su(1,1)$-algebra. 
The reason may be not necessary to mention. 
The Casimir operator ${\wtilde {\mib T}}(\sigma)^2$ for the set 
$({\wtilde T}_{\pm,0}(\sigma), \sigma=\pm)$ obeying the 
$su(1,1)$-algebra is given in the following form: 
\begin{equation}
{\wtilde {\mib T}}(\sigma)^2={\wtilde T}_0(\sigma)^2
-(1/2)({\wtilde T}_-(\sigma){\wtilde T}_+(\sigma)
+{\wtilde T}_+(\sigma){\wtilde T}_-(\sigma))
={\wtilde T}(\sigma)({\wtilde T}(\sigma)-\hbar) \ , 
\label{2}
\end{equation}
\vspace{-0.5cm}
\bsub\label{3}
\beqn
{\wtilde T}(+)&=&(\hbar/2)(-{\hat a}_+^*{\hat a}_++{\hat b}_-^*{\hat b}_-+1) 
\ , \label{3a}\qquad\qquad\qquad\qquad\qquad\qquad\qquad\qquad\\
{\wtilde T}(-)&=&(\hbar/2)(-{\hat a}_-^*{\hat a}_-+{\hat b}_+^*{\hat b}_++1) 
\ . \label{3b}
\eeqn
\esub
In the above algebra, we have four mutually commuted operators 
${\wtilde T}(+)$, ${\wtilde T}_0(+)$, ${\wtilde T}(-)$ and 
${\wtilde T}_0(-)$. 
Instead of ${\wtilde T}_0(+)$ and ${\wtilde T}_0(-)$, in this paper, 
we use 
\bsub\label{4}
\beqn
& &{\wtilde K}(+)={\wtilde T}_0(+)-{\wtilde T}(+)=\hbar{\hat a}_+^*{\hat a}_+ 
\ , \label{4a}\\
& &{\wtilde K}(-)={\wtilde T}_0(-)-{\wtilde T}(-)=\hbar{\hat a}_-^*{\hat a}_- 
\ . \label{4b}
\eeqn
\esub
The mutually commuted operators used in (A) are related to the above in the 
form 
\beqn\label{5}
& &{\wtilde L}=(1/2)({\wtilde K}(+)+{\wtilde K}(-))+({\wtilde T}(-)
-\hbar/2)\ , \qquad
{\wtilde M}=(1/2)({\wtilde K}(+)+{\wtilde K}(-))\ , 
\nonumber\\
& &{\wtilde T}={\wtilde T}(+) \ , \qquad {\wtilde K}={\wtilde K}(+) \ . 
\eeqn
Further, we have the following correspondence to the original fermion space: 
\beqn\label{6}
& &{\wtilde T}(+)\sim (\hbar/2)(\Omega_- -{\hat {\cal N}}/2+1) \ , \qquad
{\wtilde K}(+)\sim \hbar{\hat {\cal N}}_+/2 \ , \nonumber\\
& &{\wtilde T}(-)\sim (\hbar/2)(\Omega_+ -{\hat {\cal N}}/2+1) \ , \qquad
{\wtilde K}(-)\sim \hbar{\hat {\cal N}}_-/2 \ .
\eeqn
The Hamiltonian (A$\cdot$3$\cdot$3) can be expressed in the form 
\beqn\label{7}
{\wtilde H}&=&\epsilon({\wtilde T}(+)-{\wtilde T}(-))+\epsilon
({\wtilde K}(+)-{\wtilde K}(-))\nonumber\\
& &-2G({\wtilde T}(+){\wtilde K}(-)+{\wtilde T}(-){\wtilde K}(+))
-2G{\wtilde K}(+){\wtilde K}(-)\nonumber\\
& &-G({\wtilde T}_+(+){\wtilde T}_-(-)+{\wtilde T}_+(-){\wtilde T}_-(+)) \ .
\eeqn
A possible form of the orthogonal set for the above algebra is given 
as follows: 
\bsub\label{8}
\beqn
\kket{k_+,k_-;t_+,t_-}&=&\left(
\sqrt{k_+!(2t_++k_+)!k_-!(2t_-+k_-)!}\right)^{-1}\nonumber\\
& &\times ({\wtilde T}_+(+))^{k_+}({\wtilde T}_+(-))^{k_-}
({\hat b}_+^*)^{2t_+-1}({\hat b}_-^*)^{2t_--1}\kket{0} \ . 
\label{8a}
\eeqn
Here, $\hbar k_{\sigma}$ and $\hbar t_{\sigma}$ denote the eigenvalues of 
${\wtilde K}(\sigma)$ and ${\wtilde T}(\sigma)$, respectively. 
It should be noted that, in this paper, we treat the case 
\begin{equation}\label{8b}
t_+\ , \ t_-=1/2,1,3/2,\cdots ,\qquad k_+\ , \ k_-=0,1,2,\cdots \ .
\end{equation}
\esub
The state (\ref{8a}) is identical to the state (A$\cdot$7$\cdot$18) 
under the correspondence (\ref{5}).

The above is an outline of the $su(1,1)\otimes su(1,1)$-algebra in the 
Schwinger boson representation for the two-level pairing model. 
Then, it may be interesting to investigate the corresponding 
Holstein-Primakoff boson representation. 
In this case, the eigenvalue of ${\wtilde T}(+)$ and ${\wtilde T}(-)$ are 
fixed to be any of the values (\ref{8b}). 
Therefore, we can construct the representation in terms of two kinds of 
bosons, which we denote $({\hat c}_+,{\hat c}_+^*)$ and 
$({\hat c}_-,{\hat c}_-^*)$. 
The orthogonal set $\dket{k_+,k_-}$ is expressed as 
\begin{equation}\label{9}
\dket{k_+,k_-}=\left(\sqrt{k_+!k_-!}\right)^{-1}({\hat c}_+^*)^{k_+}
({\hat c}_-^*)^{k_-}\dket{0} \ . 
\end{equation}
In order to connect the state (\ref{9}) with the state (\ref{8}), 
the MYT mapping method is useful.\cite{2} 
We define the mapping operator ${\maru U}$ in the form 
\bsub\label{10}
\beqn
& &{\maru U}=\sum_{k_+,k_-=0}^{\infty}\dket{k_+,k_-}\bbra{k_+,k_-;t_+,t_-} \ ,
\label{10a}\\
& &{\maru U}{\maru U}^{\dagger}=1\ , \qquad
{\maru U}^{\dagger}{\maru U}={\wtilde P}
=\sum_{k_+,k_-}\kket{k_+,k_-;t_+,t_-}\bbra{k_+,k_-;t_+,t_-} \ .
\label{10b}
\eeqn
\esub
Then, we have 
\bsub\label{11}
\beqn
& &{\maru U}{\wtilde T}(\sigma){\maru U}^{\dagger}=T(\sigma) \ , \qquad
T(\sigma)=\hbar t_{\sigma} \ , 
\label{11a}\\
& &{\maru U}{\wtilde K}(\sigma){\maru U}^{\dagger}={\maru K}(\sigma) \ , \qquad
{\maru K}(\sigma)=\hbar {\hat c}_\sigma^*{\hat c}_{\sigma} \ . 
\label{11b}
\eeqn
\esub
Further, we have 
\beqn\label{12}
& &{\maru U}\hbar^2{\hat a}_+^*{\hat b}_-^*{\hat b}_+{\hat a}_-
{\maru U}^{\dagger}={\maru U}{\wtilde T}_+(+)
{\wtilde T}_-(-){\maru U}^{\dagger}
\nonumber\\
&=&\hbar{\hat c}_+^*\sqrt{2T(+)+\hbar {\hat c}_+^*{\hat c}_+}
\sqrt{2T(-)+\hbar {\hat c}_-^*{\hat c}_-}\ {\hat c}_- \nonumber\\
&=&\sqrt{2T(-)+\hbar{\hat c}_-^*{\hat c}_-}\cdot \hbar{\hat c}_+^*{\hat c}_-
\cdot\sqrt{2T(+)+\hbar {\hat c}_+^*{\hat c}_+} \ .
\eeqn
The Hamiltonian ${\wtilde H}$ shown in the relation (A$\cdot$3$\cdot$3) 
is mapped to ${\maru H}$ through ${\maru H}={\maru U}{\wtilde H}
{\maru U}^{\dagger}$: 
\beqn\label{13}
{\maru H}&=&
\epsilon (T(+)-T(-))+\epsilon(\hbar{\hat c}_+^*{\hat c}_+-\hbar {\hat c}_-^*
{\hat c}_-) \nonumber\\
& &-2G(T(+)\hbar{\hat c}_-^*{\hat c}_-+T(-)\hbar{\hat c}_+^*{\hat c}_+)
-2G\hbar{\hat c}_+^*{\hat c}_+\cdot \hbar{\hat c}_-^*{\hat c}_- \nonumber\\
& &-G\biggl[\sqrt{2T(-)+\hbar{\hat c}_-^*{\hat c}_-}\cdot
\hbar{\hat c}_+^*{\hat c}_-\cdot
\sqrt{2T(+)+\hbar{\hat c}_+^*{\hat c}_+}\nonumber\\
& &\qquad +\sqrt{2T(+)+\hbar{\hat c}_+^*{\hat c}_+}\cdot
\hbar{\hat c}_-^*{\hat c}_+\cdot
\sqrt{2T(-)+\hbar{\hat c}_-^*{\hat c}_-}\biggl] \ .
\eeqn
We can see that ${\maru H}$ is expressed in terms of two kinds of bosons, 
and further, it may be interesting to see that ${\maru H}$ obeys the 
$su(2)$-algebra in the Schwinger boson representation, the generators of which 
are expressed in the form 
\begin{equation}\label{14}
{\maru M}_+=\hbar{\hat c}_+^*{\hat c}_- \ , \qquad
{\maru M}_-=\hbar{\hat c}_-^*{\hat c}_+ \ , \qquad
{\maru M}_0=(\hbar/2)({\hat c}_+^*{\hat c}_+-{\hat c}_-^*{\hat c}_-) \ . 
\end{equation}
The Casimir operator ${\maru {\mib M}}^2$ is obtained as 
\beqn
& &{\maru {\mib M}}^2={\maru M}_0^2+(1/2)({\maru M}_-{\maru M}_+
+{\maru M}_+{\maru M}_-)={\maru M}({\maru M}+\hbar) \ , 
\label{15}\\
& &{\maru M}=(\hbar/2)({\hat c}_+^*{\hat c}_++{\hat c}_-^*{\hat c}_-) \ .
\label{16}
\eeqn
Clearly, ${\maru M}$ commutes with ${\maru H}$.

Next, we will investigate the squeezed coherent state, the explicit form of 
which was shown in the relation (A$\cdot$7$\cdot$20). 
This state can be written as 
\beqn\label{17}
\kket{c^0}&=&
N_{c_+}\exp\left(\frac{V_+}{\hbar U_+}{\wtilde T}_+(+)\right)\exp\left(
\frac{W_-}{\sqrt{\hbar}U_+}{\hat b}_-^*\right) \nonumber\\
& &\times N_{c_-}\exp\left(\frac{V_-}{\hbar U_-}{\wtilde T}_+(-)\right)
\exp\left(
\frac{W_+}{\sqrt{\hbar}U_-}{\hat b}_+^*\right) \kket{0} \ . 
\eeqn
Here, $W_{\sigma}$ and $V_{\sigma}$ denote complex parameters and 
$U_{\sigma}=\sqrt{1+|V_{\sigma}|^2}$, and $N_{c_{\sigma}}$ is given as 
\begin{equation}\label{18}
N_{c_{\sigma}}=(U_{\sigma})^{-1}\exp\left(-\frac{1}{2\hbar}|W_{-\sigma}|^2
\right) \ . 
\end{equation}
In this paper, instead of these parameters, we use 
$(\phi_{\sigma}, T(\sigma))$ and $(c_{\sigma},c_{\sigma}^*)$. 
We require that these new parameters obey the following condition: 
\beqn\label{19}
& &\bbra{c^0}i\hbar\partial_{\phi_{\sigma}}\kket{c^0}=T(\sigma)-\hbar/2 \ , 
\qquad \bbra{c^0}i\hbar\partial_{T({\sigma})}\kket{c^0}=0  \ , \nonumber\\
& &\bbra{c^0}\partial_{c_{\sigma}}\kket{c^0}=c_{\sigma}^*/2 \ , 
\qquad \bbra{c^0}\partial_{c_{\sigma}^*}\kket{c^0}=-c_{\sigma}/2 \ .
\eeqn
Further, we impose the condition 
\begin{equation}\label{20}
\bbra{c^0}{\wtilde T}(\sigma)\kket{c^0}=T(\sigma) \ .
\end{equation}
Under the conditions (\ref{19}) and (\ref{20}), we have the following form: 
\beqn\label{21}
& &W_{\sigma}=\sqrt{2(T(\sigma)-\hbar/2)}e^{-i\phi_{\sigma}/2} \ , \nonumber\\
& &V_{\sigma}=\sqrt{\frac{\hbar}{2T(\sigma)}}c_{\sigma} \ , \qquad
U_{\sigma}=\sqrt{1+\frac{\hbar}{2T(\sigma)}c_{\sigma}^*c_{\sigma}} \ .
\eeqn
The conditions (\ref{19}) and (\ref{20}) tell us that 
$(\phi_{\sigma},T(\sigma))$ and $(c_{\sigma},c_{\sigma}^*)$ play the role of 
the angle-action variable and canonical variable in boson type in classical 
mechanics. 
The classical Hamiltonian $H^0$ is obtained by calculating the expectation 
value $\bbra{c^0}{\wtilde H}\kket{c^0}(=H^0)$. 
The result is as follows:
\beqn\label{22}
H^0&=&\epsilon(T(+)-T(-))+\epsilon (\hbar c_+^*c_+-\hbar c_-^*c_-)
\nonumber\\
& &-2G(T(+)\hbar c_-^*c_-+T(-)\hbar c_+^*c_+)
-2G\hbar{c}_+^*{c}_+\cdot \hbar{c}_-^*{c}_- \nonumber\\
& &-G\biggl[\sqrt{2T(-)+\hbar{c}_-^*{c}_-}\cdot
\hbar{c}_+^*{c}_-\cdot
\sqrt{2T(+)+\hbar{c}_+^*{c}_+}\nonumber\\
& &\qquad +\sqrt{2T(+)+\hbar{c}_+^*{c}_+}\cdot
\hbar{c}_-^*{c}_+\cdot
\sqrt{2T(-)+\hbar{c}_-^*{c}_-}\biggl] \ .
\eeqn
We can see that the classical Hamiltonian $H^0$ corresponds 
to ${\maru H}$ shown 
in the relation (\ref{13}) under the correspondence 
\begin{equation}\label{23}
{\hat c}_{\sigma}\sim c_{\sigma} \ , 
\qquad {\hat c}_{\sigma}^*\sim c_{\sigma}^* \ . 
\end{equation}
Of course, the commutators should correspond to the Poisson bracket and 
the ordering of various terms in the Hamiltonian $H$ should obey the 
form shown in the relation (\ref{22}). 
If the order changes, $H^0$ does not become ${\maru H}$ under the 
correspondence (\ref{23}). 
The above is the Holstein-Primakoff representation of the two-level 
pairing model and its classical counterpart. 
Thus, we could learn that the squeezed coherent state (\ref{17}) with the form 
(\ref{21}) induces the classical counterpart. 

In (A), we presented a disguised form based on one kind of boson. 
Let us investigate the relation of the present disguised form to that shown 
in (A). 
For this purpose, we note the $su(2)$-algebra $({\maru M}_{\pm,0})$ shown 
in the relation (\ref{14}), together with the form (\ref{16}). 
The orthogonal set for this algebra is shown as follows: 
\bsub\label{24}
\beqn
\dket{k;m}&=&\sqrt{\frac{(2m-k)!}{k!}}\frac{1}{(2m)!}({\maru M}_+)^{k}
({\hat c}_-^*)^{2m}\dket{0} \nonumber\\
&=&\left(\sqrt{k!(2m-k)!}\right)^{-1}({\hat c}_+^*)^k
({\hat c}_-^*)^{2m-k}\dket{0} \ , 
\label{24a}\\
m&=&0,1/2,1,\cdots ,\qquad k=0,1,\cdots ,2m\ .
\label{24b}
\eeqn
\esub
The eigenvalues of ${\maru M}$ and ${\maru M}_0$ are given as 
$\hbar m$ and $\hbar (k-m)$, respectively, and of course, 
the relation to $k_+$ and $k_-$ shown in the state (\ref{9}) is 
presented as 
\begin{equation}\label{25}
k_+=k \ , \qquad k_-=2m-k \ . 
\end{equation}
Then, in the boson space constructed by one kind of boson 
$({\hat c},{\hat c}^*)$, we define the state 
\begin{equation}\label{26}
\ket{k}=\left(\sqrt{k!}\right)^{-1}({\hat c}^*)^k\ket{0} \ .
\end{equation}
This space was introduced in (A). 
The mapping operator ${\wtilde U}$ from the space 
$\{\dket{k;m}\}$ to the space $\{\ket{k}\}$\cite{2} is given as 
\bsub\label{27}
\beqn
& &{\wtilde U}=\sum_{k=0}^{2m}\ket{k}\dbra{k;m} \ , 
\label{27a}\\
& &{\wtilde U}{\wtilde U}^{\dagger}={\hat P}=\sum_{k=0}^{2m}
\ket{k}\bra{k} \ , 
\nonumber\\
& &{\wtilde U}^{\dagger}{\wtilde U}
={\hat Q}=\sum_{k=0}^{2m}\dket{k;m}\dbra{k;m}
\ . 
\label{27b}
\eeqn
\esub
Then, we have the following relations: 
\bsub\label{28}
\beqn
& &{\wtilde U}\hbar{\hat c}_+^*{\hat c}_+{\wtilde U}^{\dagger}
=\hbar{\hat c}^*{\hat c}\cdot{\hat P} \ , 
\label{28a}\\
& &{\wtilde U}\hbar{\hat c}_-^*{\hat c}_-{\wtilde U}^{\dagger}
=(M-\hbar{\hat c}^*{\hat c})\cdot{\hat P} \ , 
\label{28b}\\
& &{\wtilde U}\hbar{\hat c}_+^*{\hat c}_+\cdot \hbar{\hat c}_-^*{\hat c}_-
{\wtilde U}^{\dagger}
=\hbar{\hat c}^*{\hat c}(M-\hbar{\hat c}^*{\hat c})\cdot{\hat P} \ , 
\qquad\qquad\qquad\qquad
\label{28c}
\eeqn
\esub
\begin{eqnarray}
& &M=\hbar m \ , 
\label{29}\\
& &{\wtilde U}\sqrt{2T(-)+\hbar{\hat c}_-^*{\hat c}_-}\cdot \hbar{\hat c}_+^*
{\hat c}_-\cdot\sqrt{2T(+)+\hbar{\hat c}_+^*{\hat c}_+}{\wtilde U}^{\dagger}
\nonumber\\
&=&\sqrt{2T(-)+2M-\hbar{\hat c}^*{\hat c}}\cdot\sqrt{\hbar}{\hat c}^*
\sqrt{2M-\hbar{\hat c}^*{\hat c}}\sqrt{2T(+)+\hbar{\hat c}^*{\hat c}}\cdot
{\hat P} \nonumber\\
&=&\sqrt{\hbar}{\hat c}^*\sqrt{2T(+)+\hbar{\hat c}^*{\hat c}}
\sqrt{2M-\hbar{\hat c}^*{\hat c}}\sqrt{2T(-)+2M-\hbar-\hbar{\hat c}^*{\hat c}}
\cdot{\hat P} \ .
\label{30}
\end{eqnarray}
With the use of the relation (\ref{5}), the Hamiltonian 
${\wtilde U}{\maru H}{\wtilde U}^{\dagger}$ is reduced to the form 
(A$\cdot$4$\cdot$11). 
Therefore, we can conclude that both forms are equivalent to each other, 
and then, the merit of the use depends on the applicability. 

Finally, we will contact with the case of the classical counterpart. 
First, we introduce the variables $(\chi, M)$ and $(c,c^*)$. 
The former and the latter denote the angle and the action variable 
and the canonical variable in boson type in classical mechanics, respectively. 
Then, we set up the relation 
\begin{equation}\label{31}
\sqrt{\hbar}c_+=\sqrt{\hbar}ce^{-i\chi/2} \ , \qquad
\sqrt{\hbar}c_-=\sqrt{2M-\hbar c^*c}\ e^{-i\chi/2} \ . 
\end{equation}
We can prove that they connect with each other through the canonical 
transformation. 
By substituting the relation (\ref{31}) into the Hamiltonian (\ref{22}), 
we obtain $H^0$ expressed in terms of $(c,c^*)$ for given values 
$T(+)$, $T(-)$ and $M$. 
Concerning $H^0$ obtained in the above process, we have two comments. 
First is related to the Hamiltonian (A$\cdot$4$\cdot$14) based on the 
mixed-mode coherent state. 
We note the factor $\sqrt{2T(-)+\hbar c_-^*c_-}$ in the Hamiltonian 
(\ref{22}). 
If substituting the relation (\ref{31}) into this factor, we have 
$\sqrt{2T(-)+2M-\hbar c^*c}$. 
On the other hand, the mixed-mode coherent state in (A) gives us this factor 
in the form $\sqrt{2T(-)+2M-\hbar-\hbar c^*c}$. 
Therefore, $H$ given in the form (A$\cdot$4$\cdot$14) is different from 
the Hamiltonian obtained by $H^0$ under the replacement (\ref{31}). 
Second is related to the quantization in the form 
\begin{equation}\label{32}
c\longrightarrow {\hat c} \ , \qquad c^* \longrightarrow {\hat c}^* \ .
\end{equation}
Then, we have 
\beqn\label{33}
& &\sqrt{2T(-)+\hbar c_-^*c_-}\cdot \hbar c_+^*c_-
\sqrt{2T(+)+\hbar c_+^*c_+}\nonumber\\
&=&\sqrt{2T(-)+2M-\hbar c^* c} \sqrt{\hbar}c^*
\sqrt{2M-\hbar c^* c}\sqrt{2T(+)+\hbar c^* c} \nonumber\\
&\longrightarrow& 
\sqrt{2T(-)+2M-\hbar {\hat c}^* {\hat c}} \sqrt{\hbar}{\hat c}^*
\sqrt{2M-\hbar {\hat c}^* {\hat c}}\sqrt{2T(+)+\hbar {\hat c}^* {\hat c}}
\nonumber\\
&=&\sqrt{\hbar}{\hat c}^*\sqrt{2T(+)+\hbar {\hat c}^* {\hat c}} 
\sqrt{2M-\hbar {\hat c}^* {\hat c}}
\sqrt{2T(-)+2M-\hbar -\hbar{\hat c}^* {\hat c}} \ . 
\eeqn
The form (\ref{33}) is identical to the form (\ref{30}). 
In this sense, the ordering of various terms is important.





\end{document}